\documentclass[twocolumn,showpacs,showkeys]{revtex4}
\usepackage{epsfig}
\usepackage{amsmath}

\begin{document}
\title{Search for the Heisenberg spin glass  on rewired square 
lattices with antiferromagnetic interaction}
\author{Tasrief Surungan}
\email{tasrief@unhas.ac.id}
\author{Bansawang BJ}
\email{bansawang@science.unhas.ac.id}
\author{Dahlang Tahir}
\email{dahlang@science.unhas.ac.id}
\affiliation{Department of Physics, Hasanuddin University, Makassar, South Sulawesi 90245, Indonesia}

\begin{abstract}
Spin glass (SG) is a typical  magnetic system with frozen random spin 
orientation at low temperatures.  The system exhibits rich physical properties, 
such as infinite number of ground states, memory effect and aging phenomena.  There are two main 
ingredients considered to be pivotal for the existence of SG behavior,
namely, frustration and randomness.  For the canonical SG system,
frustration is led by the presence of competing interaction between 
ferromagnetic (FM) and antiferromagnetic (AF) couplings.  Previously, 
Bartolozzi {\it et al.} [ Phys. Rev. B{\bf 73}, 224419 (2006)],
reported the SG properties of the AF Ising spins  on scale free network (SFN). 
It is a new type of SG, different from the canonical
one which requires the presence of both FM and AF couplings.
In this new system, frustration is purely caused by the topological
factor and its randomness is related to the irregular connectvity.
Recently, Surungan {\it et. al.} [Journal of Physics: Conference Series 640, 012001 (2015)] 
reported SG bahavior of AF Heisenberg 
model on SFN.  We further investigate this type of system 
by studying an AF Heisenberg model on rewired square lattices. 
We used Replica Exchange algorithm of  Monte Carlo Method and 
calculated the SG order parameter to search for the existence of 
SG phase.
\end{abstract}

\keywords{Phase transition, Heisenberg Spin Glasses,
 Monte Carlo Simulation, Replica Exchange Algorithm}
\pacs{05.50.+q, 75.40.Mg, 05.10.Ln, 64.60.De}
\maketitle

\section{Introduction}
The study of Spin glasses has been an active research field
for almost four decades \cite{Cannella, Edwards, Kawamura, Wittmann}.
 It is a random magnetic system
which is mainly characterized by a frozen random spins
configurarion at low temperature.  This type of system 
has no total magnetization at any temperature, 
thus does not behave  like a  regular magnet.  It is put in the 
group of magnetic system due to the cooperative phenomena of the spins, 
whose low temperature phase can still be regarded as ordered phase,
i.e., a temporally ordered phase rather than spatially ordered
phase \cite{Nishimori}. The spin glass phenomenon
was  first experimentally reported in the early 
70s by Cannella and Mydosh who observed the presence of
a cusp instead of sharp peak  of ac-susceptibility \cite{Cannella}
of transition metal impurities hosted by nobel metals ($Cu$Mn and $Au$Fe).  
These are magnetic alloys with a small
amount of magnetic impurities (Mn and Fe) randomly subtituted into
the lattice of non-magnetic hosts ($Cu$ and $Au$).

It  has been well understood  that there are two main
 ingredients responsible for the existence of spin glass phase, 
namely  frustration and  randomness. A spin is frustrated  
if it can not find  satisfactory orientation in interacting 
with its neighbor spins  for  minimizing the  free energy.
This can be exemplified by spins in a unit of 
triangular lattice  with antiferromagnetic (AF) interactions; or 
by spins in a plaquette where both ferromagnetic (FM) 
and AF couplings exist. Spins interacting 
ferromagnetically (anti-ferromagnetically)
 will prefer to be aligned (anti-aligned)  to minimize
the free energy. Therefore,  for a plaquette to render
a frustration, it has to fulfill a certain
requirement, namely the product of all
existing coupling interactions has to be negative.
The FM and AF couplings are usually 
assigned by $+1$ and $-1$, respectively. 
According to this rule, a plaquette with positive product of couplings
can not lead to a frustration  as all spins are happy to each other,
in the sense that there is no conflicting orientation.

Earlier theoretical studies of spin glasses were
performed mainly on Ising systems with infinite and finite range of interaction 
\cite{Edwards,Parisi,Kawashima,Bhatt}.
Several important results were found, 
such as the replica symmetry breaking scenario, 
the existence of infinite number of pure equilibrium states and
the ultrametric trees grouping the pure states, etc.  
Some real SG materials are indeed Ising systems. 
Nevertheless, many SG systems found experimentally, including
the canonical ones, the first spin glass
system observed, are Heisenberg type where spins are three
dimensional vectors (O(3) symmetry).

There is  an existing controversy on the existence
of spin glass phase for three dimensional (3D) Heisenberg 
system. This is in contrast with the discrete
spin cases, such as the Ising model where spin glass
phase transition is observed.
For Heisenberg model, finding the the lower critical dimension $d_l$,
below which there is  no SG phase transition,  
 has  attracted a lot of interests. 
Coluzzi \cite{Coluzzi} studied  Heisenberg spins for 4D case and 
observed  SG phase  transition.
While several works \cite{Matsubara,Campos}  reported
the existence of Heisenberg SG in 3D,
 recent study by Kawamura and Nishikawa \cite{Kawamura}  reported
that  only a chiral glass exists and no 
 spin glass phase found. These various results  idicate
that  more elaborations  on the mechanism of the emergence 
of SG phenomena are required. 
 
Motivated by recent study of Heisenberg SG
 on scale free network (SFN) \cite{Tasrief14}, here we study a slightly different 
systems, namely a rewired 2D lattices with 
antiferromagnetic interaction. Both systems are 
purely anti-AF  and have irregular connectivities.
Each spin in the two systems may have different number of neighbors.
However, the connectivity structures of the
two are different as no nodes in the rewired lattice acting as a hub
like in SFN. The rewired 2D lattice
does not have small-world behaviour either, therefore
the notion of regular lattice still remains.
With respect to SG ingredients,
both systems belong to a rather new SG model
 as their  randomness and
the frustration are rendered by
factors  different from that of the canonical system.
As known, randomness in  canonical SG systems,
 is related to  the random distribution
of FM and AF couplings  and
the frustartion is due to the presence of
competing interaction between them.
This new type of model inherits  the main ingredient of SGs as
it consists of AF triangular units where spins  are frustrated.
In fact, the AF Ising model  on SFN  has been
studied, in which SG phase was observed \cite{Bartolozzi}.

The main objective of the current study is to verify 
that irregular connectivity together with frustration
induced by topological factor can also built up 
spin glass system. In addition, it is an alternative
approach for resolving the existing controversy
in 3D Heisenbern SG model. A randomly rewired 2D lattice
can be costructed  with either fully
AF interactions or mixture of AF and FM, as well
as with various cordination number.
In this study we put an extra link to each site of 
the square lattice so that the average neigbours of each 
spin is 3.0. It is  a depleted triangular lattice
which is  partially frustrated if all  couplings
are AF. The fully-frustrated (FF) system without
randomess does not have SG transition,
instead it has an oredered magnetic
phase at low temperature. An example
of this is FF clock model on triangular lattice \cite{Tasrief2004}.
In this peace of work, we will probe whether the
AF Heisenberg spins on rewired 2D lattice  exhibits 
spin glass phase transition. 
\begin{figure}
\begin{center}
\includegraphics[width=0.8\linewidth]{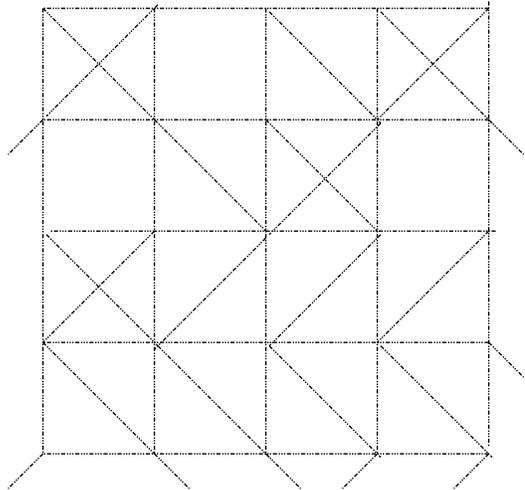}
\vspace{0.1cm}
\caption{The 2D rewired square lattice, where some sites have 
7 links, others have 6 and the rest have 5.  We consider 
peridic boundary condition.  } 
\label{Fig02}
\end{center}
\vspace{0.5cm}
\end{figure}

We use the replica exchange algorithm~\cite{hukushima96} 
of the Monte Carlo method, we calculate
the order parameters of spin glass behaviour, the so-called
overlap parameter and its distribution.  For an accurate determination
of the critical temperature,  we also evaluate the Binder parameter.
The paper is organized as follows: Section 2 describes the
model and the method. The results are discussed in
Section 3. Section 4 is devoted for a summary and  concluding
remarks.

\section{Model and Method of Simulation} 
The Heisenberg model on a rewired 2D lattices  can be written
with the following Hamiltonian,
\begin{equation}
H = \sum_{\langle ij \rangle } \vec s_i \cdot \vec s_j
\end{equation}
where $\vec s_i$ and $\vec s_j$ are 3D vector spins, each  
 occupying a  site  of the lattice.
The summation is performed  over  all directly connected spins.
In a regular lattice such as the square lattice, all
spins have the same number of neighbours, associated to
the integer coordination number. Here, we add one extra link
to each site and randomly connect it to one of its next-nearest
neighbors. Therefore, there are some sites having  seven  links, 
other having six and the rest only five. One particular realization
of the lattice is shown in Fig. \ref{Fig02}.

By excluding the double 
counting, the average number of links (ANOL) is 2.5. The lattice
system can be regarded as  a quasi regular lattice and in 
principle we can generate a lattice system with any fractional ANOL.  
As shown in the figure, the lattice consists of a large number
of  triangular units.
Spins on each triangular unit are frustrated. 
Therefore, adding extra links corresponds to increasing  the degree
of frustration. In principle, it is possible to define 
the degree of frustration and study its contribution
to the properties of the system, in particular
the SG behaviour.

We performed Monte Carlo simulation using the replica exchange algorithm 
\cite{hukushima96} to calculate thermal averages of the physical quantities of 
interest.  This algorithm is particularly well
 to overcome the slow dynamics due to the existence of local minima 
in the energy landscape. The slow dynamics is a common problem 
in dealing with complex systems such as
SGs where the random walker tend to be trapped at certain 
local minimum. It is an extended Metropolis algorithm where M replicas are
calculated in parallel. Each replica is  in equilibrium with a heat bath of inverse temperature. 
Given a set of the inverse temperatures, $\beta$, the probability distribution of
finding the whole system in a state 
 $\{ X \} =
\{ X_1,X_2, \dots, X_M\}$ 
is given by,
\begin{equation}
 P(\{ X, \beta\}) = \prod_{m=1}^{M} \tilde{P}(X_{m},\beta_{m}),
\end{equation}
with
\begin{equation}
\tilde{P}( X_m, \beta_m) = Z(\beta_{m})^{-1} \exp(-\beta_{m} H(X_{m})),
\label{equil}
\end{equation}
and $Z(\beta_m)$ is the partition function for the m-$th$ replica.
We can then define an exchange matrix between  replicas,
 $W(X_m,\beta_m| X_n,\beta_n)$, which is the probability
to switch the configuration $X_m$ at the temperature $\beta_m$
with the configuration $X_n$ at $\beta_n$. 
With the requirement to keep the entire system at equilibrium,
we use the detailed balance condition on the transition matrix
\begin{eqnarray}
P(\{ X_m, \beta_m \},\ldots, \{ X_n, \beta_n \},\ldots )\cdot
 W(X_m,\beta_m| X_n,\beta_n) \nonumber \\ = P(\{ X_n, \beta_m \},
\ldots, \{ X_m, \beta_n \},\ldots )
\cdot W( X_n,\beta_m | X_m,\beta_n), 
\end{eqnarray}
along with Eq.~(\ref{equil}), so that we have 
\begin{equation}
\frac{ W( X_m,\beta_m | X_n,\beta_n)}{ W( X_n,\beta_m | X_m,\beta_n)}=\exp(-\Delta),
\end{equation}
where $\Delta=(\beta_{n}-\beta_{m})(H(X_{m})-H(X_{n}))$.
With this constraint, we can choose the matrix coefficients
according to the standard Metropolis method which gives
the following
\begin{equation}
W(X_m,\beta_m| X_{n},\beta_n)=\left \{  \begin{array}{ccc} 
1 & {\rm if} & \Delta<0, \\ 
\exp(-\Delta) & {\rm if} & \Delta>0.
\end{array} \right.
\label{trans}
\end{equation}
Due to the fact that  the acceptance ratio decays exponentially 
with $(\beta_n-\beta_m)$, we restrict the exchange temperatures next to
each other, i.e.,   the terms $W(X_m,\beta_m|
X_{m+1},\beta_{m+1})$. 
The replica exchange algorithm has been widely 
 implemented in the study of various spin glass systems,
including the  AF-SFN Ising model where spin glass observed. 
In the next section, we present the results of our study.

\section{Results and Discussion}
\subsection{Energy and the specific heat }
We have simulated AF Heisenberg model on rewired square 
lattice of several systems with linear sizes L= 32, 48, and 64.  
We implemented periodic boundary condition
so that each site of the native square lattice has 
four neighbours.  For each system size, we took many realizations 
of the lattice, then average the results over the number of realizations.
This is a standard procedure in probing random systems such as 
SG.  Each realization corresponds to one
particular connectivity distribution which is randomly
generated. For the results to be reliable,
we have to take large number of realizations. 
Previous study of Heisenberg SG on SFN  we took 1000 realizations. 
Here We took smaller number of realizations (N=250) due to its degree
of randomness  is less compared to the previous system.

To check the reliability of our simulation
we evaluated  the energy time series of the system, from which we can 
obtain the average energy and specific heat. One
can tell whether the system is in equilibrium or not
by analysing the energy time series.
In Monte Carlo simulation, time corresponds to a series of MCSs.
 One MCS is defined as one loop of  updating each spin
in the lattice, based on its probability.
We perform $M$ 
MCSs for each temperature and take $N$ samples out of $M$. 
To make sure the system is well equilibrated we perform
enough initial MCSs, usually 10 MCSs, before doing measurement.  

\begin{figure}
\begin{center}
\includegraphics[width=1.0\linewidth]{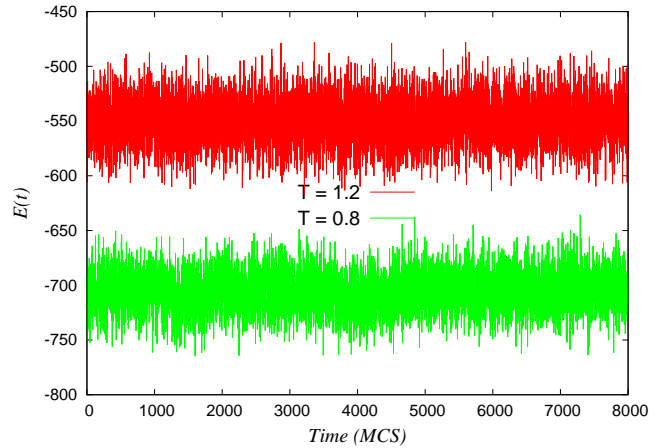}
\caption{Time series of energy for 
linear size L=32 at temperature $T=1.2$ and $T=0.8$.}
\label{TS32}
\end{center}
\end{figure}

\begin{figure}
\begin{center}
\includegraphics[width=1.0\linewidth]{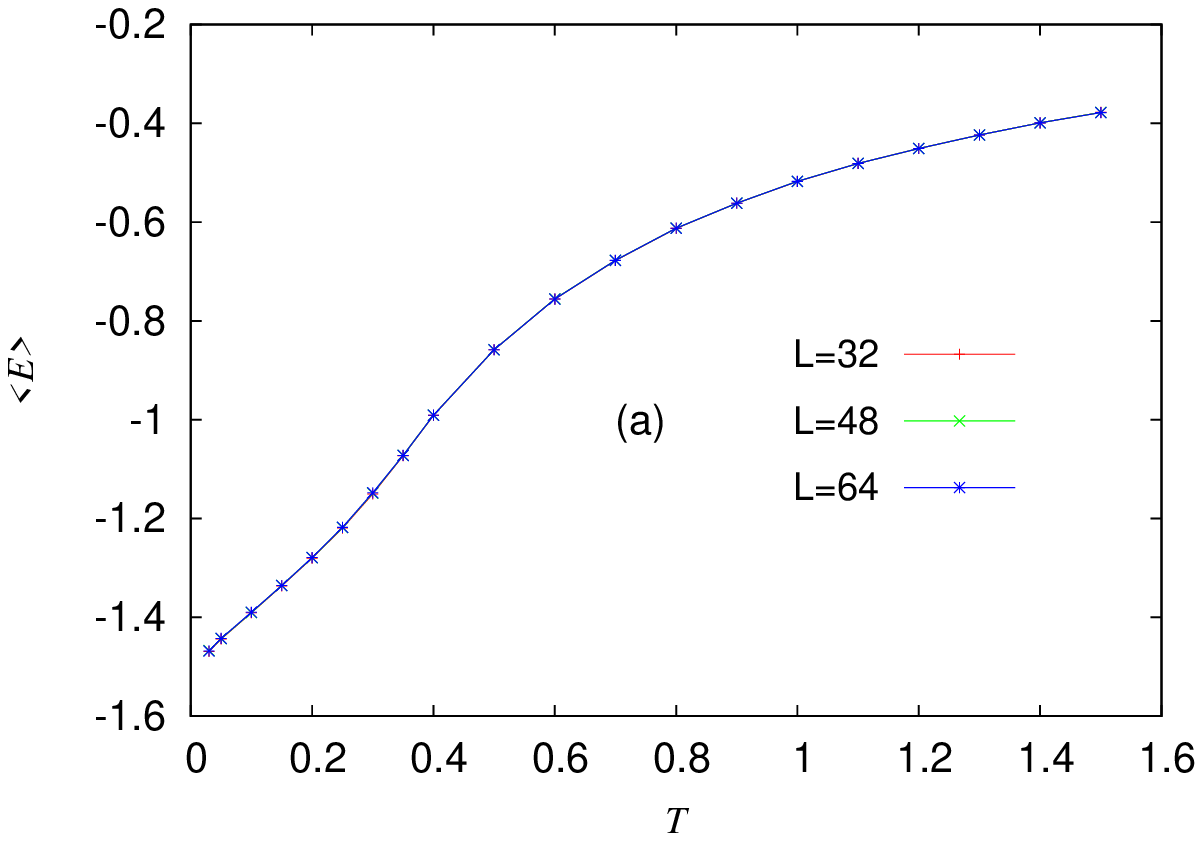}
\includegraphics[width=1.0\linewidth]{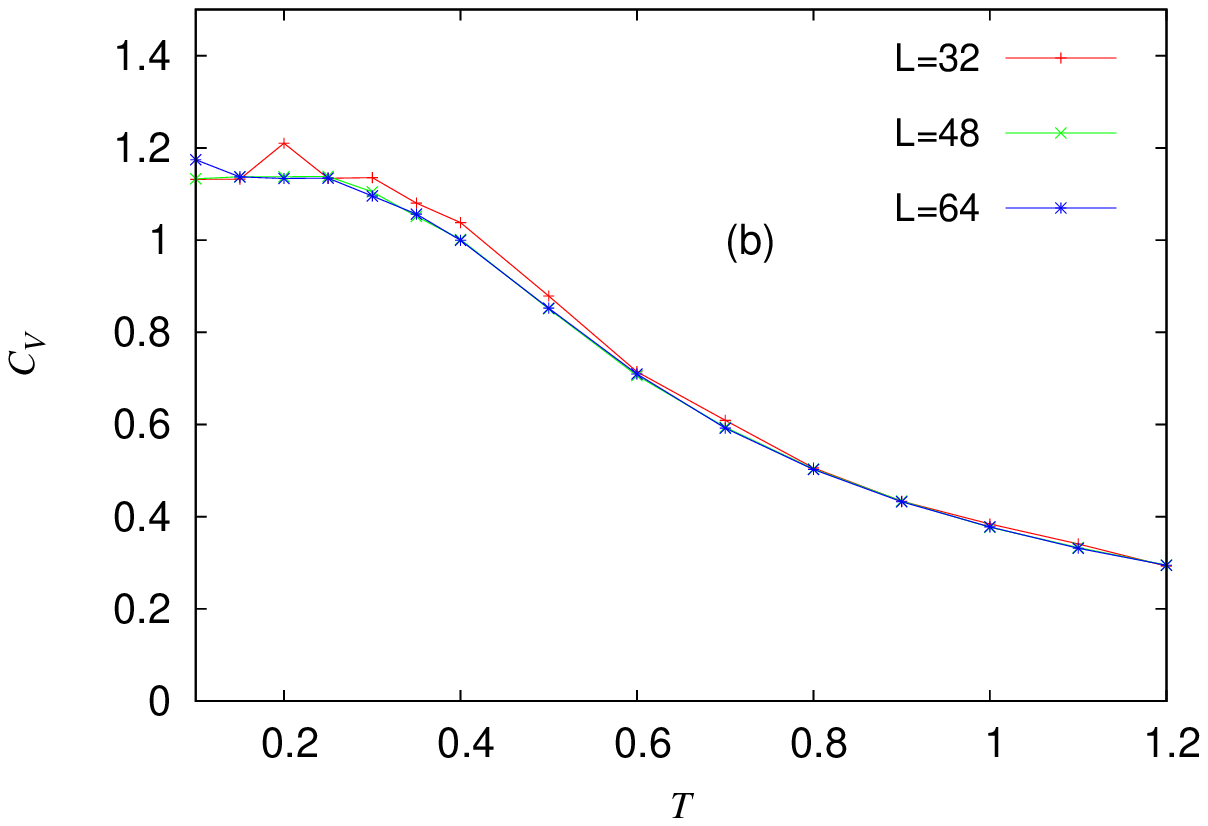}
\vspace{0.1cm}
\caption{Temperature dependence of Average energy (a) and specific heat (b)
of three different sizes, $L=32, 48 $ and $64$.}
\label{espht}
\end{center}
\vspace{0.5cm}
\end{figure}

The time series plot of energy for two different 
temperatures, i.e., T=1.2 and 0.8, for linear size $L = 32$ is shown in Fig. 
\ref{TS32}. 
This plot indicates that the system is well  equilibrated
after enough initial MCS, here it was taken 6000 MCSs.
The fluctuation at higher temperature is larger due to 
 larger thermal fluctuation. 
%
%
We extracted two quanties from energy time series,
namely the ensemble average of energy, 
$\langle E \rangle =  \frac{1}{N} \sum_N E_i$, and
the specific heat which is defined as follows
\begin{equation}
C_v = \frac{N}{kT^2} \left( \langle E^2  \rangle  - \langle E \rangle^2\right)
\end{equation}
where $N$ and $k$ are respectively the number of spins
and Boltzmann constant. The plot of these quantities are shown 
in Fig. \ref{espht}. The specific heat plot has no a clear peak 
at finite temperature, which may signify that the SG
transition is found only at $T=0$. To clarify this
we calculate the SG order parameter which is 
presented in the next sub-section.

\subsection{Spin Glass Order Parameter}
To search for the exitence of SG phase transition, 
we calculate the overlapping parameter, also
called as SG order parameter defined as follows
\begin{equation}
q_{EA} =  \langle \left|\sum_i \vec s_i^{\alpha} \otimes 
\vec s_i^{\beta} \right| \rangle_{av}
\end{equation}
This quantity basically originated from the scalar 
product of the vector spins.
A scalar product of two vectors
will give maximum value if they are parallel.
If system is frozen, their overlapping
parameter will give finite value.
In the language of Ising spin, each term of this 
quantity is the overlapping of
 two possible states (up or down).
For Heisenberg case, the overlaping
of two spins is the dot product two 3D vectors..
Because we have to accomodate the
condition where vector spins
may rotate in any direction, we take the tensor product
instead of dot product, resulting in
nine components of the quantity.
\begin{figure}
\begin{center}
\includegraphics[width=1.0\linewidth]{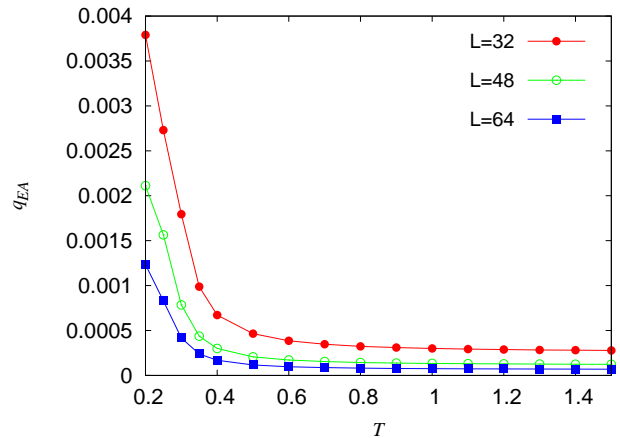}
\vspace{0.1cm}
\caption{Temperature dependence of SG order parameter 
of four different sizes, $L=32, 48 $ and $64$.}
\label{qea}
\end{center}
\vspace{0.5cm}
\end{figure}

The plot of temperature dependence of $q_{EA}$ for three
different sizes, $L = 32, 48$ and  $64$ is shown in
Fig. \ref{qea}. As indicated, the order parameter is increasing
as temperature decreases, e.g., for $L=32$, $q_{EA} > 0$ at $T < 0.4$. 
 However, as system sizes become
larger, the value of $q_{EA}$ tend to decrease.
This is an indication that the parameter goes 
to zero at infinite value of $L$, which is a sign
of the absence of SG phase at the thermodynamic limit.
The result is consistent with the plot of specific
heat which exhibits that SG phase transition
occurs at zero temperature, in other words
no finite temperature SG phase observed.
The study of the model for various connectivity
densities is still in progress. The results will
be reported elsewhere.

\section{Summary and Conclusion}
In summary, we have studied the AF Heisenberg spins on 
rewired square  lattices and searched for the existence of SG phase. One extra link
is added to each site of the lattices; which  randomly
connects the site to one of its next-nearets neighbors.
The system is  randomly frusrated due to the existence
of abundance of triangular units. It is a new type of
SG model which inherits the main ingredients
of SGs, i.e., randomness and frustration.
By using Replica Exchange Monte Carlo method,
which is a standard method in SG study, we calculated several physical quantities, such
ensemble average of energy, the specific heat and the overlappig parameter.
We observed no finite temperature spin glass phase transition.
This result suggested that the the AF rewired 2D lattice
with average connection 3.0 can not host the existence of SG phase.
System with such average connectivity may correspond
to the canonical models below the critical dimension.
Further consideration of rewired 2D lattice 
with larger connectivity densities is required.

\section*{Acknowledgments}

The authors wish to thank K. Hukushima and  M. Troyer for valuable discussions.
One of the authors (TS) is grateful to  the hospitality of Theoretical
Physics Division of CERN during his academic visit to the Center
and to the Indonesian Ministry of Research and Technology for
supporting the visit. The computation of this work was performed
using parallel computing facility in the Department of Physics
Hasanuddin University and the HPC facility of Indonesian 
Institute of Science. The work is supported by HIKOM (Hibah Kompetensi) 
research grant 2014 of the Indonesian Ministry of Education and Culture. 



\begin{thebibliography}{9}
\bibitem{Cannella} V. Cannella and J. A. Mydosh, Phys. Rev. B, {\bf 6},4220 (1972)
\bibitem{Edwards}S. F. Edwards and P. W. Anderson, J. Phys. F 5, 965, (1975).
\bibitem{Kawamura} H. Kawamura and S. Nishikawa, Phys. Rev. B {\bf 85}, 134439 (2012).
\bibitem{Wittmann} M. Wittmann and A. P. Young, Phys. Rev. E {\bf 85}, 041104, (2012) 
\bibitem{Nishimori}H. Nishimori, {\it Statistical Physics of Spin Glasses and Information 
Processing}, Oxford Univ. Press, (2001).
\bibitem{Parisi} G. Parisi, Phys. Rev. Lett. 50, 1946, (1983).
\bibitem{Bhatt}R. N. Bhatt and P. Young Phys.Rev. B. 54, 924, (1985)
\bibitem{Kawashima}N. Kawashima and A. P. Young, Phys. Rev. B {\bf 53}, R484, (1996).
\bibitem{Coluzzi}B. Coluzzi, J. Phys. A: Math. Gen. 28 , 747 (1995).
\bibitem{Matsubara}F. Matsubara1, T. Shirakura, S. Endoh and S. Takahashi,
 J. Phys. A: Math. Gen. 36 10881, (2003).
\bibitem{Campos}I. Campos {\it et al}, Phys. Rev. Lett. 97, 217204, (2006)
\bibitem{Tasrief14}T. Surungan, F. P. Zen,  and A.G. Williams,  
Journal of Physics: Conference Series, 640, 012001 (2015). 
\bibitem{Bartolozzi} M. Bartolozzi, T. Surungan, D.B. Leinweber and 
   A.G. Williams,  Phys. Rev. B{\bf 73}, 224419 (2006).
\bibitem{Tasrief2004} T. Surungan, Y. Okabe, and Y. Tomita, J. Phys. A 37, 4219, (2004).
\bibitem{hukushima96} K. Hukushima and K. Nemoto, J. Phys. Soc.
Japan {\bf 65}, 1863 (1996).
\end{thebibliography}
\end{document}